\RequirePackage{fix-cm}
\documentclass[smallextended]{svjour3}       
\smartqed  
\usepackage{graphicx}
\usepackage{caption}
\usepackage{algorithm}
\usepackage{algorithmic}
\usepackage{amsmath}
\usepackage{balance}
\usepackage{amssymb}

\usepackage{colortbl}
\usepackage{booktabs}
\usepackage{multicol}

\usepackage{multirow}
\begin{document}

\title{$P^4QS$: A Peer to Peer Privacy Preserving Query Service for Location-Based Mobile Applications
}


\author{Meysam Ghaffari         \and
        Nasser Ghadiri	\and
        Mohammad Hossein Manshaei	\and
        Mehran Sadeghi Lahijani
}


\institute{  \at
              Department of Electrical and Computer Engineering, Isfahan University of Technology, Isfahan 84156-83111, Iran.
              \email{\{meysam.ghafari,m.sadeghi\}@ec.iut.ac.ir  \{nghadiri,manshaei\}@cc.iut.ac.ir}        
           \and
           \at
             Corresponding Author: Nasser Ghadiri
}

\date{Received: date / Accepted: date}

\maketitle

\begin{abstract}
The location-based services provide an interesting combination of cyber and physical worlds. However, they can also threaten the users' privacy. Existing privacy preserving protocols require trusted nodes, with serious security and computational bottlenecks. In this paper, we propose a novel distributed anonymizing protocol based on peer-to-peer architecture. Each mobile node is responsible for anonymizing a specific zone. The mobile nodes collaborate in anonymizing their queries, without the need not get access to any information about each other. In the proposed protocol, each request will be sent with a randomly chosen ticket. The encrypted response produced by the server is sent to a particular mobile node (called broker node) over the network, based on the hash value of this ticket. The user will query the broker to get the response. All parts of the messages are encrypted except the fields required for the anonymizer and the broker. This will secure the packet exchange over the P2P network. The proposed protocol was implemented and tested successfully, and the experimental results showed that it could be deployed efficiently to achieve user privacy in location-based services.

\end{abstract}

\section{Introduction}
With the technological growth, especially in the Internet and communication devices, available services are increasing. A new class of services that have emerged with the development of the Internet and smartphones with location sensors are location-based services (LBS). An LBS system provides one or more location-dependent services customized for every user that connects to the LBS provider and sends her location information to the server.

As LBS combines the cyber and physical worlds, it would be interesting for users. Example applications are finding nearby places such as restaurants and gas stations, or finding nearby friends through location-based friend finders, such as Foursquare.

Although such services are attractive, they need access to physical information about the real-world user's life \cite{almuhimedi2015your}. If the system fails to provide adequate security mechanisms, it can be harmful to the users' privacy. For example, an intruder can infer the user's behavior, habits, social activities, sickness and other information just by knowing her location~\cite{gutwirth2002privacy}. Therefore, the users' security and privacy   must be regarded as one of the most important issues in LBS.

The privacy aspects of the location data have been extensively investigated. Most researchers acknowledge the need to hide the users' locations even from the service providers \cite{hengartner2007hiding}. With the development of storage devices and the increased processing power of computers, the users could be easily identified by tracking their movements and the locations that they pass through. Furthermore, analyzing such trajectory data is possible \cite{lin2014mining}, and most LBS users are not comfortable with this issue. It can cause many threats to the users' privacy. For example, users may be exposed to location advertisements, or subject to prejudice by service providers, or  physical harm and extortion \cite{gutwirth2002privacy}. So, many researchers have proposed different methods such as spatial and temporal cloaking \cite{gruteser2003anonymous}\cite{ying2015protecting}, fake locations \cite{you2007protecting}, the addition of noise \cite{hoh2005protecting}, \textit{k}-anonymity \cite{gruteser2003anonymous}, and similar methods to cope with this threat.

A well-known identity protection mechanism is \textit{k}-anonymity \cite{kalnis2007preventing}. The principle operation of this method is as follows: The system will preserve the privacy of a location-dependent query from a user if the probability of distinguishing the user's query is less than $\frac{1}{\textit{k}}$. A trusted third-party anonymizer is used to achieve this goal \cite{gedik2008protecting}\cite{gruteser2003anonymous}\cite{kalnis2007preventing} \cite{mokbel2006new} (see Fig.~\ref{fig:TrustServer}). The trusted server interacts with users and manipulates their location data. Users will send their queries to the anonymizer instead of sending them directly to the LBS server. The anonymizer generates a cloaking region around the user in a way that the user is indistinguishable from \textit{k}-1 other users. Although this method seems to be working, it has its shortcomings. A serious challenge is to find a tradeoff between privacy, time and the quality of service.

A familiar privacy metric is based on the concept of anonymity, implying that the user is indistinguishable from other users as members of an anonymity set. Another privacy metric is unlinkability, which means the attacker cannot distinguish whether two or more items of interest have a relationship between them. In unlinkability approach, the information revealed by a sequence of queries is considered important because of the computation and storage characteristics. Most of the acquired information items are gained by analyzing the sequence of queries \cite{pfitzmann2010terminology}. We will employ these metrics to preserve anonymity and unlinkability.

As discussed above, most of the existing methods are just a trade-off between the quality of service and privacy. Some of them suffer from the risk of revealing information in security bottlenecks, leading to sensitive information about the users to be disclosed. While the unlinkability factor is crucial, it has not been considered thoroughly in literature. 

In this paper, a novel architecture named Peer to Peer Privacy Preserving Query Service ($P^4QS$) is proposed for improving the users' privacy in location-based social networks. First, we use a distributed P2P architecture to acquire the queries and deliver the results. Second, by using proper symmetric and asymmetric encryption methods, each node would have access to the necessary information only. Third, by exploiting the Distributed Hash Table (DHT) in the proposed P2P architecture, the response is delivered to the client, but the LBS could not identify him. The main contributions of this paper are threefold:
\begin{enumerate}
\item We propose a novel P2P model for anonymizing the location-dependent queries. In this model, the nodes collaborate with each other to make the queries \textit{K}-anonymized. At the same time, the information contents of the queries are protected.

\item The protocol has four advantages: 
\begin{itemize}
\item It does not require any trusted nodes. There are few distributed methods for anonymization, but they need a few trusted nodes to achieve anonymity.
\item Unlike existing distributed methods, the nodes do not need to store any information such as geographical maps.
\item The system has no security, privacy or computational bottleneck. For example in central trusted server, the server itself is the bottleneck of the service.

\item There is no additional or hidden costs for deploying the system. Unlike existing methods that would be expensive to implement due to their need for specific hardware and systems, the proposed method could be easily implemented with no prerequisites.
\end{itemize}
\item We have implemented the $P^4QS$ as a proof-of-concept prototype to evaluate its feasibility. Full source code is made available for  download\footnote{http://dkr.iut.ac.ir/content/codes-Peer-to-Peer-Privacy-Preserving-Query-Service}.
\end{enumerate}

The rest of the paper is organized as follows: in Section~\ref{Related Works}, the related works will be surveyed. In Section~\ref{Proposed Model}, the basic concepts and a high-level view of the proposed model will be introduced. In Section~\ref{Protocol Description}, the proposed model and protocol description will be explained and evaluated in detail. In Section~\ref{Evaluation}, both analytical and numerical evaluations of the protocol are illustrated by using a set of common scenarios and experiments. Finally, Section~\ref{Cunclusion} closes the paper with the conclusion.

\section{Related Works} \label{Related Works}
\label{Sec:Relatedwork}
In this section, we give an overview of the current literature in location privacy domain. The techniques are categorized to the pseudonym, perturbation, and trusted server methods, as well as distributed architectures explained in the following subsections.

\subsection{Pseudonym}
Using the real identity has the risk of revealing information about users. The pseudonym method is proposed to address this problem. It uses a fictitious name instead of the user identifier \cite{beresford2003location}. However, it has been proved that by tracing the sequence of activities and queries, the identity of the user is detectable. The mix zone method further develops this core approach, which defines specific zones and allows the pseudonyms to change when the user changes his region  \cite{lu2012pseudonym} \cite{beresford2004mix}. By changing the pseudonym, the unlinkability metric is satisfied whenever the user moves into another area.

\subsection{Perturbation and Obfuscation}
Another class of methods to achieve location privacy use perturbation and obfuscation. Adding noise, spatial cloaking, and fake location methods fall into this category.

The noise addition method interleaves noisy queries between the user queries to hide the original query. A problem with this approach is the high computational overhead on the server to process all queries \cite{hoh2005protecting}. The fake location method improves this approach by sending fake queries. Besides the real path of the user and his queries, a false path of queries will be generated and sent to hide the actual path of the user. However, the adversary could distinguish the authentic and fake paths by using background knowledge, map matching, or other methods such as using spatiotemporal density estimation and line intersections\cite{hoh2005protecting}\cite{bhattacharya2015automatically}.

Spatial cloaking is a general method based on concealing the user queries by adding cloaks to achieve the \textit{k}-anonymity. In this way, the user will become indistinguishable from \textit{k}-1 other users. In the primary method, a constant cloaking range is added to user queries. This method is optimized by proposing the trusted anonymizer server as explained in the following subsection.

\subsection{Trusted Server}
\begin{figure}[t]
\centering
\includegraphics[scale=0.25]{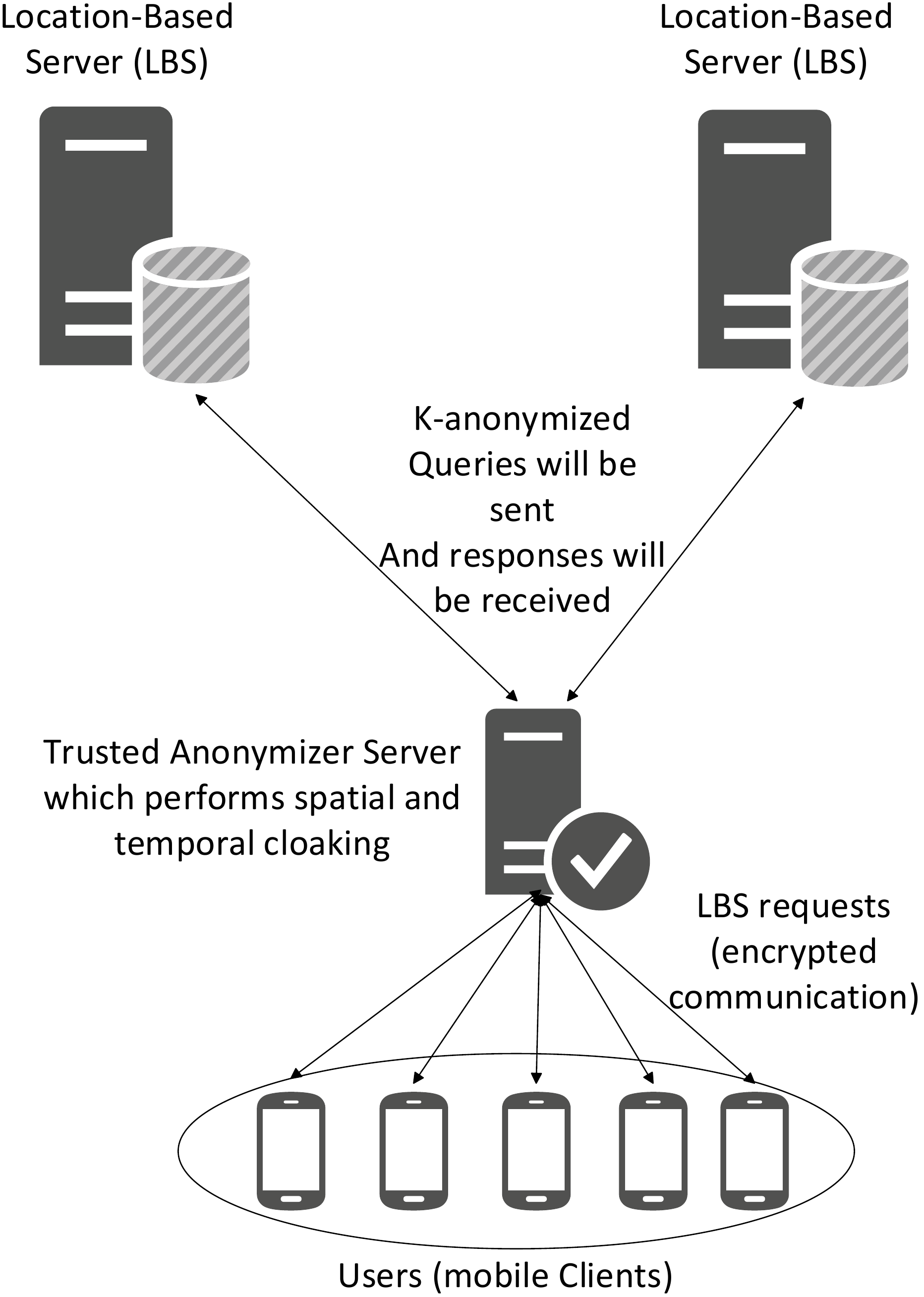}
\caption{Central Trusted Server architecture. In this architecture, all nodes send their requests to one trusted server. This server makes the queries K-anonymized and sends them to the LBS.}
\label{fig:TrustServer}
\end{figure}

The methods above such as spatial cloaking, fake location, and noise addition put significant computational overhead on the server or imply the loss of quality. To overcome this problem, the use of trusted anonymizer server became popular, making a balance between privacy and the quality loss. The general architecture of the trusted server is shown in Fig.~\ref{fig:TrustServer}. The basic idea is that each node sends its query to the trusted server. The trusted server accumulates the queries and anonymizes them by adding spatial and temporal cloaking in a way that at least \textit{k} different queries become indistinguishable.

 The trusted server architecture was first proposed by Gruteser and Grunwald \cite{gruteser2003anonymous}. It provides \textit{k}-anonymity for protecting privacy based on a trusted anonymization server. They represent location information of the client by three tuples $((x_1,y_1),(x_2,y_2),(t_1,t_2))$. Tuples one and two show the area and the third tuple indicates the  period of client queries. Based on this structure, they design a cloaking algorithm that generates spatiotemporal cloaking. The cloaked region contains, at least, \textit{k} clients and their location information are sent to the location-based server indistinguishably. The parameter \textit{k} is the minimum acceptable anonymity. In this method, it is very tough to provide a flexible privacy protection scheme, because of its pre-defined constant value \textit{k}. Moreover, in the case of small areas, \textit{k}-anonymity is very hard to achieve.

Significant efforts have been made to overcome these drawbacks and some novel methods are proposed, including CliqueCloak \cite{gedik2008protecting}. CliqueCloak is a personalized anonymity method by which clients can tune their personal privacy protection requirements as well as their spatiotemporal cloaking levels. This personal level is achieved by modeling the anonymization constraints to find the best satisfaction conditions. Another framework proposed for customized anonymity is Casper \cite{mokbel2006new}. Casper has a profile for each user's privacy settings to fulfill the requirements of every individual user. It consists of a value that determines \textit{k} for the minimum acceptable cloaking in \textit{k}-anonymity. Casper also uses an incomplete pyramid structure to keep the user information dynamically \cite{aref1990efficient}. 
 
 Another scheme based on trusted server is Cache Cloak \cite{meyerowitz2009hiding}. This method caches LBS responses to use them for subsequent queries. Thus, it decreases the number of queries and the chance of adversary to infer from queries. Exploiting this method can increase privacy without the loss of accuracy. However, this system suffers from scalability problem. It cannot be used in a large scale situation because it needs huge amounts of memory to save the queries and responses. Moreover, in the worst case, if each client has  different types of queries from different location-based servers, the trusted server needs to cache an enormous amount of information. Thus, it becomes the bottleneck of their proposed system.

Schlegel et al. proposed dynamic grid system (DGS) which requires a semi-trusted third party to preserve user privacy~\cite{schlegel2015user}. In this method, the user could define her privacy level. In DGS different privacy levels has not different communication costs for the user. However, this method also needs a central server.  
 
In all derivations of central trusted anonymizer server, the queries must be sent to the central anonymizer. This limitation makes the anonymizer a bottleneck for the system. If an intruder penetrates to the trusted server, he could extract client's information. It has been proved that using pseudonyms or even spatial cloaking cannot solve this problem~\cite{ghaffari2014ambiguity}.

We propose a novel architecture based on distributed trusted server to address the problems mentioned above. In our proposed protocol, even an intruder can attack, only a limited information would be revealed. The proposed method relies on a peer-to-peer distributed architecture. Before presenting our solution, we review some literature in this category.

\begin{table*}[ht]
\caption{Functions of each peer in the system.}
\begin{center}
\begin{tabular}{l | p{7cm}} \toprule
{ \textbf{ Party}}  & { \centering{\textbf{Function}}}   \\  \rowcolor[gray]{.9} \hline
$P_{C_i}$: \emph{Client Peer i}  & Sends a query and waits for the response.  The system must defend against correlation of queries and $P_{C_i}$.\\
$P_{A_j}$ :\emph{Anonymizing Peer}  & Each user node determines a user's chosen location and is responsible for the anonymization of queries  \\
& in the zone nearby. If sufficient anonymization is not achieved, this node cooperates with nearby $P_{A_j}$s. \\ \rowcolor[gray]{.9}
 $P_{B_m}$: \emph{Broker Peer}  & Receives the response packet based on \emph{Hash(Ticket)} and 
       waits for $P_{C_i}$ to ask for his packet. \\\rowcolor[gray]{.9}
&Then $P_{B_m}$ sends the response packet to $P_{C_i}$.\\
\hline
\end{tabular}
\end{center}
\label{tbl:Roles}
\end{table*}

\subsection{Mobile and Distributed Architectures}
Central trusted servers had the potential of becoming the single point of failure, causing system failure. The central servers also have the risk of being compromised or controlled by an intruder, making them not reliable. Thus, deploying a secure trusted server that works smoothly in real-world situations would be extremely hard to achieve. To solve this problem, mobile-based methods have been proposed. The advantage of the mobile-based method for the users is that they could preserve privacy without relying on the central servers. In this approach, several groups of mobile users collaborate with each other to achieve the desired privacy.

CAP provides \textit{k}-anonymity by maintaining road density using quadtree followed by VHC mapping and perturbation \cite{pingley2009cap}. VHC mapping is used to project a two-dimensional geographic space into a one-dimensional space which preserves the homogeneity of every point. The adjacent points will also remain close to each other \cite{shin2012privacy}. In this way CAP tries to reduce the storage and computational overheads required for mobile-based methods of storing the maps to achieve \textit{k}-anonymity.

Hu and Xu proposed a method to generate cloaking boxes based on the strength of receiving a signal from the nearby devices \cite{hu2009non}.

Chen proposed the LISA method. LISA is based on \textit{m}-unobservability and tries to prevent the attacker from attributing any particular location to the user. LISA also predicts the user movements by exploiting Kalman Filter that leads to increased privacy \cite{chen2009energy}. 

Our proposed protocol employs mobile-based methods that are used to preserve anonymity and unlinkability. Unlike the methods mentioned above, the privacy of user will be maintained without revealing and exchanging location information. Users communicate with each other, but they cannot get any information about other user's location or queries. Unlinkability metric  is also satisfied by using ticket instead of user identification number, the pseudonym, or any other identifier.
 
\section{Proposed Model} \label{Proposed Model}

The proposed method which is called Peer to Peer Privacy Preserving Query Service ($P^4QS$) follows a role-based, peer-to-peer architecture. As it is shown in Table~\ref{tbl:Roles}, each node may have different roles at any time to perform one or more of three functionalities: (1) querying, (2) anonymizing queries, and (3) brokering responses.

%
%
%

From an overall view, the system operates as follows. In the first step, a node needs to send its query. We call this node "\emph{Client Peer}" because it needs the service and plays the role of a client in a peer-to-peer architecture. We designate clients by $P_{C_i}$, where C $\in$ \{1,2,...,N\}. Each user is located in a zone where she sends the query. We represent these zones by $Z_{i}$, where $i$ shows the sum of latitude and longitude of the user. 

The $P_{C_i}$'s query will be sent to another peer responsible for anonymizing the specific zone which contains the query location. The anonymization is the second functionality of each peer. In this role, the "\emph{Anonymizing Peer}" gets queries from different clients corresponding to a specific zone for which this Anonymizing Peer is in charge.  We  represent this peer by $P_{A_j}$. Note that there exist $N$ Anonymizing Peers, and each peer is responsible for the zone $Z_{i}$, which is represented  by $P_{A}^{Z_{i}}$. 

\begin{figure}[h]
\centering
\includegraphics[scale=0.28]{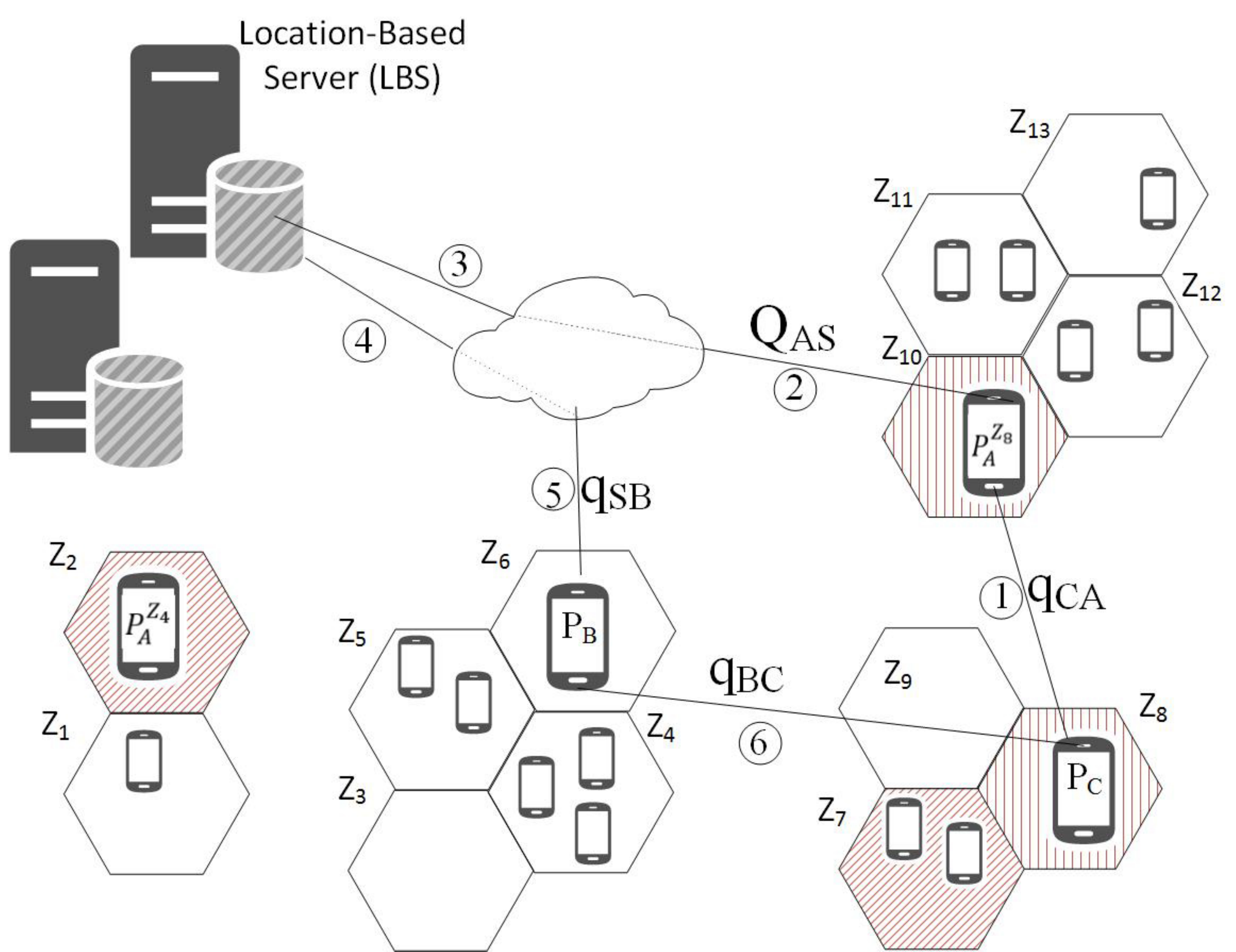}
\caption{$P^4QS$ Architecture: Anonymizing and sending a query to the distributed system and getting the response. \textcircled{1} Client Peer sends query $q_{CA}$ which includes Location, Proposed key, Ticket and Query. The proposed key is encrypted with the server’s public key using RSA method. Ticket and query are encrypted with the proposed key using AES encryption method. \textcircled{2},\textcircled{3} The Anonymizing Peer adds spatial and temporal cloaking to the queries and sends $Q_{AS}$ set of K-anonymized queries to the LBS. \textcircled{4} The LBS sends the encrypted (response) $q_{SB}$ with the clients proposed key to the appropriate broker. \textcircled{5} The broker will be chosen based on the calculated \emph{Hash(Ticket)}. \textcircled{6} The client interacts the broker and gets its response $q_{BC}$. }
\label{fig:model}

\end{figure}

$P_{A_j}$ anonymizes the queries from multiple $P_{C_i}$s by adding appropriate spatial and temporal cloaking to them to make them K-anonymized and sends them to the LBS. The LBS processes the queries and sends each of them to the specific peer responsible for that response. This peer, which is called \emph{Broker Peer}, is represented by $P_{B_m}$, where $B$ $\in$ \{1,...M\}. The Broker Peer gets the responses from the LBS and keeps them until the specific client asks its response from the broker.  The broker then sends the corresponding answer to its owner, i.e., each $P_C$ asks the response from $P_{B_m}$ and $P_{B_m}$ sends back the response to the $P_{C_i}$. The overall process is shown in Fig.~\ref{fig:model}.

An essential point in the proposed architecture is that no central trusted server is required. Thus, the risk of the single point of failure will be eliminated. As shown in Algorithm~\ref{alg:Pseudomodel}, we first choose the nearest $P_{A_j}$ based on the nearest neighbor of the query location (i.e., Line $3$ in Algorithm~\ref{alg:Pseudomodel}). Achieving K-anonymity needs \textit{K} different queries. So $P_{A_j}$ needs to wait for queries from other users, or it has to generate fake queries. There is a wait time loop (line $4$) such that if $P_{A_j}$ does not receive enough queries (line $5$), it will collaborate with the adjacent nodes or generate fake queries, as shown in line $6$ of the algorithm. Then \textit{K}-anonymized queries are sent to the server. The server will process the queries and send the encrypted response packet to $P_{B_m}$ (i.e., line $11$). $P_{B_m}$ is chosen based on the \emph{Hash (Ticket)}, where the ticket is sent by Client Peer in the first step. Finally, $P_{C_i}$ asks \textit{R} from $P_{B_m}$ and $P_{B_m}$ sends it to $P_{C_i}$. 

 So the overall $P^4QS$ protocol procedure is as follows: $P_{C_i}$ finds the appropriate $P_{A_j}$ and also finds $P_{B_m}$ based on Hash(Ticket). Then $P_{C_i}$ sends its query to the $P_{A_j}$ and request for the response of this query to the $P_{B_m}$. The $P_{A_j}$ accumulates and anonymizes the queries. If needed, $P_{A_j}$ adds fake queries to achieve the predefined privacy threshold. Then it sends these anonymized queries to the server. The server gets the queries,  decrypts and processes each query, and sends the response to the node $P_{B_m}$ based on \emph{Hash(Ticket)}. After receiving a response from server or a request for response from a client, $P_{B_m}$ matches the request and response and forwards the answer to the client. The overall process was shown in Algorithm~\ref{alg:Pseudomodel}. The user's query is in the form of:  
 \begin{small}
 $$Loc, E( prop-key)_{Pub_{server}}, E(Ticket, Query, P_{B_m}'s IP)_{prop-key} $$
 \end{small}
 where Loc is the location of the user and the query is based on this place.
 
 \begin{enumerate}
 \item The random number which is generated by the LBS to determine $P_{B_m}$.
 \item A validation time stamp which shows the expiry time of the ticket.
 \item The server mark which is encrypted with the private key of the server. Every user can decrypt it with the public key of the server to assure the validity of the ticket, in the case of receiving a ticket from other peers.
 \end{enumerate}    
       
        $\textit{prop-key}$ is the user's proposed key for the encryption of the response. This key can be changed for each request. $Pub_{server}$ is the server's public key and is known for all users.
     
    \begin{algorithm}
      \caption{The Pseudo code of the $P^4QS$}
    \begin{algorithmic}[1]
     \label{alg:Pseudomodel}
   \STATE $ \textbf{Input:} Client Query: $\\ Encrypted with proposed key(Ticket, Query) and Encrypted with server’s public key(proposed key)
   \STATE $ \textbf{Output:} Response $\space$ of $\space$ the $\space$ query $\space$ which $\space$ will $\space$ be $\space$ delivered $\space$ to $\space$ Client $\space$ with $\space$ a $\space$ distributed $\space$ hash $\space$ table$
    \STATE $P_{C_i} $\space$ Sends $\space$  Query $\space$ to $\space$ P_{A_j} $
     \WHILE {Anonymity in $P_{A_j}$ $<$ K}
    \STATE $ Wait $\space$ for $\space$ new $\space$ query$
    \STATE $ Or$
    \STATE $Collaborate $\space$ with $\space$ P_{A_j} \pm 1 $\space$ (Adjacent $\space$ P_{A_j})$ or $\space$ generate $\space$ fake $\space$ queries
     \ENDWHILE
    \STATE $P_{A_j}  $\space$ sends $\space$ (K $\space$ Anonymized $\space$ Query)$\space$  to $\space$ the $\space$ server $
    \STATE $Server $\space$ process $\space$ queries $
    \STATE $	Server $\space$ sends $\space$ E(R)_{Proposed Key} $\space$ to $\space$ P_{B_m}  $\space$ which $\space$ P_{B_m}= Hash(Ticket) $
    \STATE $P_{C_i}$  $\space$ Sends  $\space$ Query  $\space$ to  $\space$ $P_{A_j}$
    \STATE $P_{B_m} $\space$ Sends $\space$ R $\space$ to $\space$ P_{C_i} $
    
    \end{algorithmic}
    \end{algorithm}

To evaluate the efficiency of the proposed model, we use a set of common challenging scenarios to assess this model from multiple aspects. These scenarios have different frequency of queries, time delay sensitivity, position accuracy and the amount of revealed information per query. Consequently, each scenario can evaluate the model from a different aspects. The features of these scenarios and their attributes are summarized in Table~\ref{tbl:Scenarios}. We will explain the scenarios and evaluate the model by them in the evaluation section.

 \begin{table*}[ht]
\begin{center}
   \captionof{table}{Proposed scenarios and their usages}
    
      \begin{tabular}{   p{4cm} | p{3cm} | p{3cm}  }\toprule
         
      \centering{\textbf{Scenario}} & \centering{\textbf{Main feature}} & \textbf{Proven feature by Scenario}\\  \rowcolor[gray]{.9} 
      
       \hline
        
       \emph{Profile Matching (PM)}: used for finding similar profiles & Accuracy is important but sequence of queries must not correlated & preserving accuracy, simultaneously no correlation between queries 
          \\ 
          
       \hline
      \emph{Driving Condition Monitoring (DCM)}: in case of need to monitor the road condition, privacy of the users who sent the data must be protected & low sensitivity to the position accuracy and delay but sequence of queries must be kept uncorrelated & correlation between queries in case of not sensitivity to delay and position accuracy  \\
      \rowcolor[gray]{.9} 
       \hline
      \emph{Road Map (RM)}: used to find the path & Sensitive to delay but Position cloaking is tolerable & ability to protect against path tracking \\
       \hline
      \emph{Detecting User By Sequence of Queries (DUSQ)}: can be used to detect user and visited places & Revealing user's identity by correlating queries to user & inability to associate queries to user \\\rowcolor[gray]{.9} 
        \hline
        \emph{Detecting User Speed in Highway (DUSW)}: even without detecting the user identity, detecting over-speeding in highways can jeopardize privacy& inability to correlate sequence of queries to each other & detecting correlation between queries \\
          \hline           
      \end{tabular}
 \label{tbl:Scenarios}
  \end{center}
   \end{table*}

\section{Protocol description}  
\label{Protocol Description}

The $P^4QS$ is based on attaining two important goals: distributed anonymizing and avoid correlation between the sequences of queries (unlinkability). To achieve these goals, we proposed a distributed peer-to-peer architecture. In the proposed system, each peer has three roles, as described in the previous section, namely \emph{client}, \emph{anonymizer}, and \emph{broker} roles.

In the first step, each peer needs to send its query, but, unlike the previous methods, the query will be forwarded to the relevant anonymizer peer. Each peer in our system is responsible for anonymizing queries close to a particular location. The anonymizer peer gets multiple queries from multiple client peers. These queries have an important part in common, and that is their positions. These locations of queries are close to each other and $P_{A_j}$ anonymizes them and sends them to the LBS.
 
 The structure of query is slightly different from typical queries. The query consists of the user's location which is sent as plain text, the ticket, the query text and the proposed key. The proposed key is encrypted with the public key of the LBS and the ticket and the query encrypted with the proposed key. The ticket is used as a control mechanism in our protocol and will be described later in this section. The queries are encrypted with the user's proposed key. For peer-to-peer communications, we exploit a distributed hash table (DHT) as described shortly. Then we will explain the proposed method and show how it uses the DHT.
 
 \subsection{Distributed Hash Table (DHT)}

\begin{figure}[ht]
 \centering
 \includegraphics[scale=0.30]{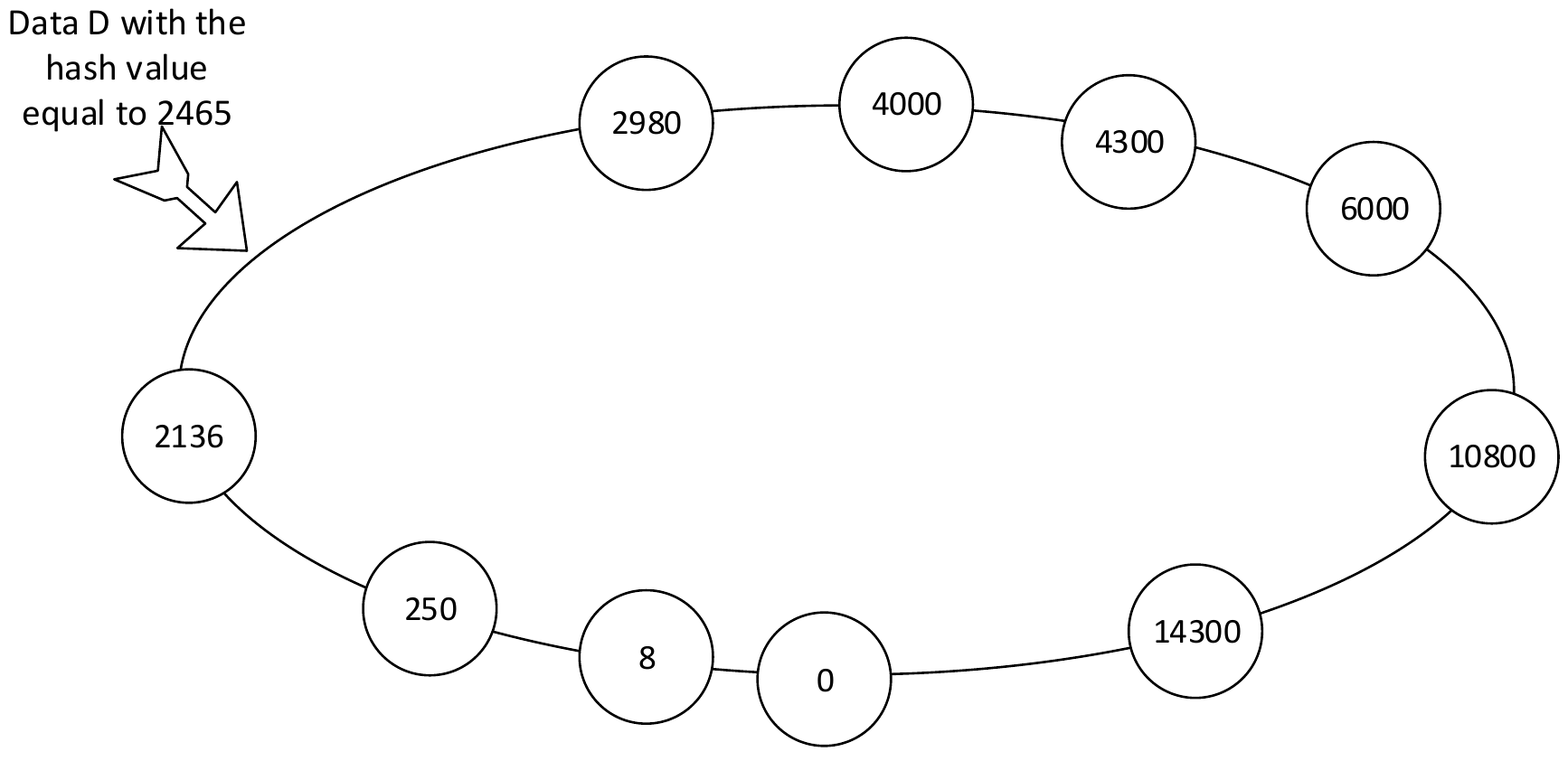}
 \caption{DHT Search Space: each node is specified by an index number from 0 to $2^n-1$. For finding a specific node, the user finds the closest node to the \emph{Hash(data)}. In this case, there exist $10$ active peers. The hash value of $D$ is equal to $2465$. Hence, peer $2136$ is responsible for it.}
 \label{fig:DHT}
 \end{figure}

 Distributed hash table belongs to a class of decentralized systems for distributing tasks between peers. In this method, each peer is responsible for keeping data and forwarding it to others. The DHT architectures provide three major features: scalability, flexibility, and immediate deployment. It could be used efficiently in mobile nodes \cite{ribe2014hierarchical}. For finding a specific node in the DHT, different methods have been proposed. One of the best and most simple methods is Chord. In Chord, each peer stores the addresses of $n$ peers. The computational complexity of the search operation is $log(N)$, where $N$ is the number of all peers in the distributed table \cite{stoica2001chord}. As depicted in Fig.~\ref{fig:DHT}, if a peer has data and wants to locate or save this data, it will calculate the hash of data. The nearest peer to the extracted value is responsible for keeping or locating the data. For example, in Fig.~\ref{fig:DHT}, peer X is responsible for data "D" as its hashed value is the closest to the hash of data. Most of the proposed DHT use SHA-1 hash function.
 
 \subsection{Hashing Protocol}
One of the advantages of DHTs is that the user could specify the appropriate hash function. In the proposed system, the protocol uses geographical data for this purpose, which consists of latitude and longitude. A straightforward and efficient method for using these two-dimensional data is to map it into a new one-dimensional space. For this purpose, we define our hashing function by, 
 
 \begin{equation}
 H=X+Y,
 \end{equation}
where X and Y are the longitude and latitude values of the determined location of $P_{C_i}$ and H is the hash value used in DHT for finding a particular peer. Using this function, we can simply find the best anonymizer peer for each location. Adjacent geographical anonymizer peers (peers whose RAs are close to each other) are located beside each other in the DHT, so they can easily cooperate with each other to achieve the best anonymization by optimum spatial and temporal cloaking.  
 
\begin{figure}[ht]
 \centering
 \includegraphics[scale=0.25]{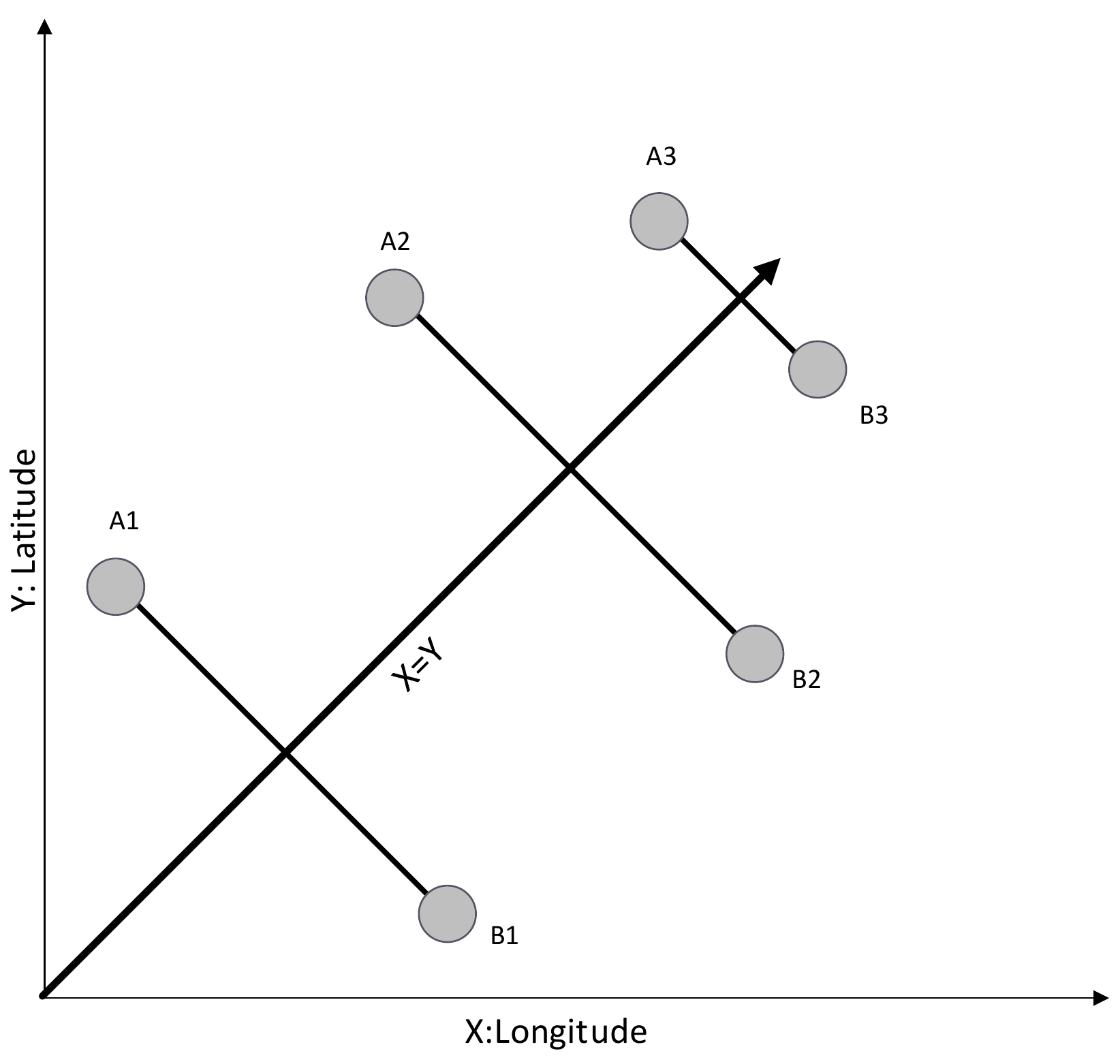}
 \caption{The symmetric points. If some nodes locate in each of these symmetric locations, they will be assigned to one anonymizer based on the sum of their latitude and longitude.}
 \label{fig:xandy}
 \end{figure}

 The proposed function has one weakness in the first glance: the symmetric points will be matched to one point in the new space, as shown in Fig.~\ref{fig:xandy}.
 
 This problem could be addressed by assigning two symmetric zones to each anonymizer. Thus, each anonymizer is responsible for both regions. After receiving the queries, the anonymizer separates them based on a defined threshold and anonymizes each zone separately.    
 
 \subsection{Ticket}  \label{Ticket}
 A key architectural element in our proposed method builds on the concept of the ticket. It will be used for running the $P^4QS$ protocol by different peers. Unlike pseudonym-based methods, exchanging the ticket does not need any computation or authentication. Therefore, it could be easily implemented and quickly used. Moreover, analyzing linkability between queries is impossible. The usage of the ticket is described in this subsection.  
 
  When the location-based server processes the query and produces the response, it needs to send it back to the client. The aim of our protocol is to make the client completely anonymous and eliminate the linkability of the queries to other queries or users. An identifier is needed to identify the owner of each query and simultaneously change it abruptly with the minimum computation cost. For this purpose, we propose using a unique ticket for each query. By this method, the owner of the query is unknown to the LBS and the only available information is the ticket, which is unique for each query. Thus, analyzing a sequence of queries becomes impossible. 
 
  The server(LBS) sends the response based on the ticket. A simple method for the LBS is extracting the hash of the ticket and sending the response to the appropriate peer. The peer with the closest hash to the hash of the ticket, which is called \emph{Broker Peer}, is responsible for receiving the response from the server (LBS). In this case, the same hash function could be used in this step as well, and there is no need to define a new distributed hash table.
  
 The \emph{Client Peer} knows the ticket and thus, it can calculate the hash of the ticket and find out the appropriate \emph{Broker Peer} to ask for the response.

 \subsection{Protocol Execution}  \label{Protocol Execution}
 The $P^4QS$ protocol executes in two phases: initialization and execution. In the initialization phase, the peers introduce themselves to the system. In the initialization process, each peer runs a server thread and listens to a particular port. Then it chooses \textit{myRA}, which is a random number used for determining the anonymizer and the broker.
  
 After receiving the response to the join request, the client is ready to perform all his functions as $P_{C_i}$ , $P_{A_j}$ and $P_{B_m}$ . The pseudo codes for these functions are shown in Algorithm~\ref{alg:PC}, Algorithm~\ref{alg:PA}, and Algorithm~\ref{alg:PB}, respectively.
  
    \begin{algorithm}
    \caption{Pseudo code of $P_{C_i}$ (\emph{Client Peer})}
    \begin{algorithmic}[1]
    \label{alg:PC}
    \STATE $Get $\space$ the $\space$ Location$
    \STATE $Choose $\space$ myRA $
    \STATE $Calculate $\space$ Hash=X+Y$  
    $of $\space$ the $\space$ Location $  
    \STATE $Choose $\space$ one $\space$ of $\space$ the $\space$ tickets $\space$ and $\space$ retrieve $\space$ its $\space$ Hash $\space$ by decrypting $\space$ the $\space$ ticket $\space$ with $\space$ Server’s $\space$ public $\space$ key$
    \STATE $Find $\space$ Anonymizer$
    \STATE $Find $\space$ Broker$ (Based $\space$ on $\space$  ticket)
     \STATE $Send $\space$ Query:$\space$ Loc,E(PropKey)_{Pub_{Server}}, $\space$ E(Ticket, query, Broker's IP address)_{PropKey} $
    \STATE $Send  $\space$ "Request-For-Answer" $\space$ message $\space$ to $\space$ Broker$         
   
    \end{algorithmic}
    \end{algorithm}
    
   As shown in Algorithm~\ref{alg:PC}, $P_{C_i}$ finds the appropriate anonymizer based on $Hash(location)$ and sends the query to it. Also, the broker is chosen based on the $Hash(Ticket)$. So, by changing the users' location, the anonymizer peer will be changed. Moreover, each response is sent to a broker. Thus, compromising a client (a broker peer) by the server will not reveal significant information about a specific user.  
    
     \begin{algorithm}[h]
       \caption{The Pseudo code of $P_{A_j}$ (\emph{Anonymizing Peer})}
     \begin{algorithmic}[1]
     \label{alg:PA}
     \WHILE {Anonymizer Peer is Active}
      \WHILE {Anonymity in $P_{A_j}$ $<$ K}
      \IF {$New $\space$ Query $\space$ Recieved$}
     \STATE $Put $\space$ query $\space$ in $\space$ ReceivedQueries $\space$ list $ 
     \ENDIF
     \STATE $ Wait $\space$ T $\space$ second $\space$ for $\space$ new $\space$ queries$
     \IF{ $ after $\space$ T $\space$ $\space$ second $\space$ Queries<K $}
     \STATE $Collaborate $\space$ with $\space$ P_{A_j} \pm 1 $\space$ (Adjacent $\space$ P_{A_j})$
     \STATE $\textbf{OR} $
     \STATE $Generate $\space$ Fake $\space$ Query(ies) $
     \ENDIF     
      \ENDWHILE
      \STATE $P_{A_j}  $\space$ sends $\space$ (K $\space$ Anonymized $\space$ Queries)$\space$  to $\space$ the $\space$ server $
      \STATE $Empty $\space$ Query $\space$ List$     
      \ENDWHILE    
     
     \end{algorithmic}
     \end{algorithm}
     
Algorithm~\ref{alg:PA} shows the function of the anonymizer peer. In this procedure, the anonymizer is always waiting for new queries. In lines 3 and 4, after a new query is received, the anonymizer puts it in the query list. Lines 6-11 indicate that the anonymizer waits for \textit{T} seconds to gather \textit{K} queries. If the number of received queries is less than \textit{K}, she  can collaborate with the adjacent anonymizers or create some queries herself. Then, the anonymizer adds spatial cloaking to the K queries to make them indistinguishable. Since the ticket information is encrypted by the public key of the LBS, the anonymizer has no access to the sensitive information of the user, so she just could anonymize the user location. Thus even in case of having a malicious or semi-honest anonymizer she just has access to the user location in a specific zone. As soon as the user moves to an adjacent zone, she will be assigned to another anonymizer so the anonymizer could not analyze the user path.
     
     Functioning as a broker peer, the client saves requests for the answers received from other clients and the responses received from the LBS in two separate lists. As soon as a new request for answer or response is received, it checks the list and if it finds a match, it will forward the response to the $P_{C_i}$ and remove both request and response from the lists (as shown in Algorithm~\ref{alg:PB}).

\begin{algorithm}[h]
           \caption{The Pseudo code of $P_{B_m}$ (\emph{Broker Peer})}
         \begin{algorithmic}[1]
         \label{alg:PB}
         \WHILE {Broker Peer is Active}
         \STATE $Wait $\space$ for $\space$ new $\space$ message$         
          \IF {$New $\space$ Response (Request $\space$ for $\space$ Response) $\space$ Received$}
          \IF {$There$\space$ is $\space$ a $\space$ Match $\space$ for $\space$ it $\space$ in $\space$ Request$\space$ List(Response $\space$ List) $}
         \STATE $Remove $\space$ the $\space$ Matched $\space$ Request $\space$ (Response) $\space$ from $\space$ the $\space$ Request $\space$ List $\space$ (Response $\space$ List)  $ 
         \ENDIF
         \ELSE
         \STATE $PUT $\space$ the $\space$ Request $\space$ (Response) $\space$ in $\space$ the $\space$ Request $\space$ List $\space$ (Response $\space$ List)$
         \ENDIF
                
\ENDWHILE 
\end{algorithmic}
\end{algorithm}    


As described in Section~\ref{Ticket}, the system uses tickets to deliver responses to the owner.  Although using multiple pseudonyms is an ideal method to achieve privacy, it has the risk of the (Denial of Service) DoS attack, leading to system breakdown \cite{avoine2013privacy}. The $P^4QS$ is similar to multiple pseudonyms method, but it does not require the pseudonym changing process, eliminating the need for time and computational process. The $P^4QS$ could be vulnerable to DoS attacks especially because, unlike multiple pseudonyms methods, it does not have any authentication. So, the proposed method is ideal for privacy, because it uses tickets that work like a pseudonym for each query. However, the protocol needs to be protected against DoS attacks. To solve this problem, we propose a ticket exchange protocol.
    
    If an intruder wants to make a DoS attack on the system, he will need to send plenty of valid queries. We prevent such attacks by limiting the allowed number of queries of each user. To achieve this goal, we propose a ticket exchange protocol. In this protocol, the location-based server generates a collection of tickets and sends a specific number of tickets to each authorized user of the system in predefined time periods; for example, in an hourly manner. The client uses one of these tickets for each query. Thus, the DoS attack becomes preventable. However, this ticket exchange approach has two issues: using randomly generated tickets by the attacker and tracking the users by knowing the owner of the tickets. To solve these problems, we further improve the ticket exchange protocol.
 
\textbf{Ticket Generation:} 

The server will generate a pre-defined number of tickets and send them to the users in a specific time periods. The structure of ticket sent by the LBS is $E(Token,Identifier, Valid Time)_{Priv_{Server}}$, which means the ticket is an encrypted package including a random token, an identifier and the valid time of the ticket. This package is encrypted with the private key of the LBS. So each entity in the system can decrypt it with the public key of the LBS and check the identifier to control the validity of the package. This mechanism has two important roles in the system. First, the attacker could not generate random tickets and the LBS checks the identifier. Before processing the query, the LBS checks the validity of the ticket number and thus, the chance of the intruder will be lowered significantly. The second and more important role of this identifier is that in the $P^4QS$, users exchange some of their tickets with each other in defined time intervals frequently. The tickets are chosen randomly for exchange and each user exchanges tickets just with the adjacent peers. The user checks the identifier after getting each ticket  and thus, she can validate the tickets. If any user gets any invalid tickets, she drops them. If the invalid tickets exceed a predefined threshold, she will consider the sender as untrusted, so generating and distributing fake tickets will be prevented. 

After a few exchanges, the tickets will be circulated across the network of the users, and the LBS may not be able to determine the owner of any query. Thus, we prevent DoS attack and provide the ideal privacy. The pseudo code for the ticket exchange protocol is shown with Algorithm~\ref{alg:Ticket exchange}. In this algorithm, $P_{C_{t_i}}$ is a list of trusted peers for peer $ P_{C_i} $ and $ P_{C_{t_k}} $ is a peer in this list.    
    
      \begin{algorithm}
       \caption{Ticket Exchange Protocol}
       \begin{algorithmic}[1]
       \label{alg:Ticket exchange}
       \STATE $\textbf{Ticket: } (token, verifier, Time Stamp)_{private_{server}} $
       \STATE $ Server $\space$ sends $\space$ ticket $\space$ batch $\space$ $\space$ to $\space$ each $\space$  authenticated $\space$ user $\space$ in $\space$ specific $\space$ time $\space$ periods$ 
       \WHILE{$ not $\space $ achieved  $\space $ ideally $\space $ exchanged  $\space $ tickets $}
       
       \STATE $ P_{C_i} $\space $  $\space $ exchange $\space $  E $\space $ of $\space $ its $\space $ randomly $\space $ selected $\space $ Tickets $\space $ with $\space $ other P_{C_{j}}  $
       \STATE $ Check  $\space $ the   $\space $ validity $\space $  of  $\space $ each  $\space $ received  $\space $ ticket $
       \IF{$Received $\space $ ticket $\space $ is $\space $ invalid$}
       \STATE $Drop $\space $ the $\space $ ticket $
       \STATE $ count $\space $ number $\space $ of $\space $ invalid $\space $ tickets $\space $ from $\space $ specific $\space $ P_{C_{j}} $
       \IF {$ P_{C_{j}} $\space $ sends $\space $ invalid $\space $ tickets $\space $ more $\space $ than $\space $ threshold $}
       \STATE $Remove $\space $ P_{C_{t_k}} $\space $ from $\space $ trusted $\space $ list $
       
       \ENDIF
       
       \ENDIF
       \ENDWHILE
       \STATE $ Each $\space $ peer $\space $ has $\space $ limited $\space $ valid $\space $ tickets $\space $ that $\space $ the $\space $ server $\space $ could $\space $ not $\space $ trace $\space $ them$
       \end{algorithmic}
       \end{algorithm}

   In our ticket exchange protocol, if the LBS has \textit{T} tickets for each peer in the defined periods, and the system has \textit{N} peers, every peer will exchange \textit{E} tickets in each round. The exchange could be done by sending \textit{E} tickets for the following peer. So after \textit{R} rounds, the probability of having tickets assigned by the LBS to each peer can be calculated by Equation~(\ref{eq:tickets}).
    
    \begin{equation}\label{eq:tickets}
     P=\Big(\frac{T-E}{T}\Big)^R+ \Big(\frac{E}{T}\Big)^N \sum_{i=1}^{R-1}\Big(\frac{T-E}{T}\Big)^i
     \end{equation}
    
 If  the number of peers gets high enough, as compared to the number of tickets, The sigma part of the equation could be neglected. Achieving low \textit{P} is essential for the privacy of users (preferably $ P < \frac{1}{K}$).
  
\subsection{Handling Overloading Anonymizer}
Every system faces inappropriate situations. Most of the system failures and attacks are due to these problems. In this case, discovering the weaknesses and handling the problem is critical. Based on our analysis, in the case of using the $P^4QS$, the overloading anonymizer exceptions need to be handled properly. As described in the protocol description method, each anonymizer peer is responsible for anonymizing specific zones. As these zones are randomly selected for each peer, it is possible to have an anonymizer in a crowded zone. In this case, it is possible to have too many queries. The anonymizer is just a mobile node, and it is unable to handle all queries. In this case, a right  solution is to allow this node to leave the DHT and rejoin with a new random \textit{MyRA}. The anonymizer is now responsible for the new zones. At the same time, when the anonymizer leaves the DHT, the queries will be sent to the closest peers, which means that the queries will be divided between two peers, the predecessor and successor nodes.  
 
\section{Evaluation} \label{Evaluation}
In this section, we evaluate the $P^4QS$ by providing common scenarios in Section~\ref{Scenarios} and this is followed by  evaluation against the scenarios in Section~\ref{Scenarioevaluation}. A proof-of-concept prototype of the system is also built and evaluated in Section~\ref{experiments}.

\subsection {Scenarios} \label{Scenarios}

We illustrate the different aspects of the proposed method through a few scenarios. These services are different from the following aspects:
\begin{itemize}
\item \emph{The frequency of access}: This defines the acceptable period between two sequential queries.
\item \emph{Time accuracy and delay sensitivity}: The adequate temporal cloaking which may be caused by the anonymizer due to network delay.
\item \emph{Position accuracy}: The acceptable location cloaking which is made by the anonymizer.
\item \emph{The amount of required (revealed) data}: In the case of requesting a specific service, what information must be sent (revealed) by the user?
\end{itemize}

\subsubsection{Profile Matching}
One of the possible applications of location services could be profile matching. In this case, a user needs to know specific information about users nearby. It is very important just to reveal the permitted information, and the user should not be able to perform further analysis by sending multiple queries. In this case, accuracy is critical, but a sequence of queries or responses must not be correlated.  

\subsubsection{Driving Conditions Monitoring}
Modern vehicles have multiple sensors to acquire road and weather conditions. These sensors could be useful for reporting road conditions instead of using multiple expensive fixed sensors. The vehicle will report road condition alongside its path, and the whole road will be covered with sensors. In this case, the position accuracy is not so critical, and a distance of 100 meters will be sufficient \cite{gruteser2003anonymous}. This situation is not sensitive to delay, and reporting on the condition with some delay is tolerable. However, the method must prevent the extraction of further information such as path and speed from the reported conditions \cite{wang2015privacy}\cite{yu2016senspeed}.

\subsubsection{Road Map}
The roadmap is a useful feature in routing. In this case, the user needs to report his position and query the map and specific points of interest such as hospitals, hotels, and so on. In this case, a quick and accurate response is needed, but the query could have some spatial cloaking because the answer will cover a large area around the query. The important issue here is that the identity or goal of the user must not be revealed from sequences of queries or a particular location.

\subsubsection{Detecting User based on the Sequence of Queries}
 It is important that the intruder must not be able to reveal the user or his goal by analyzing sequences of queries, regardless of the type of service or user's  intention. In this case, two important issues must be noticed. First, there is no correlation between queries or between the queries and users. Second, the exact position of the user in any query must not be revealed, because in this case, it is probable that intruder could identify the user and his goal is based on the query and background knowledge. For example, if a user queries hospitals in midnight, when he is at his home, the intruder finds that the specific person is sick, so she is going to reach a hospital.

\subsubsection{Detecting User Speed in Highways}
Sometimes just knowing the sequence of queries even without knowing the user could be harmful to the user. For example, the police or insurance companies could detect over-speeding in highways and detect the user in cameras based on his position. In this case, the location-based service is not desired, and the user prefers not to use it. Thus spatial cloaking or time delay is not a solution here, and the only possible solution is that there would be no way to correlate the sequence of queries. By doing that, it is probable that next query could be from another user, and the privacy of the user might be assured.  

\subsection{Scenarios Evaluation} 
\label{Scenarioevaluation}

In this section, we are going to analyze the $P^4QS$. First, we use the mentioned scenarios to evaluate and analyze it thoroughly from different aspects. 

The most frequent scenarios could be driving condition monitoring and road maps. In these scenarios, the accuracy and delay are ignorable. The aim is to hide the users' identity in these scenarios, as it could be achieved in most of the previous methods such as fake locations, noisy queries, trusted server, and so on. The same goal could be obtained with the $P^4QS$ because the user's query will be anonymized by spatial and temporal cloaking in the anonymizer peers.  
 
In the scenario of detecting user by a sequence of queries, the intruder needs to know the owner of multiple queries. This is possible in the previous methods such as pseudonyms. In the $P^4QS$, each query has its unique identifier, which is the ticket, and thus the attacker will not be able to analyze the sequence of queries. This situation happens for the profile matching scenario by which the attacker knows the user and wants to distinguish it from other users, thus tracing his activities. In this kind of attack, using fake paths does not solve the problem because the attacker can easily distinguish them by knowing the exact pattern of the user. But the proposed model is resistant against this attack also by using tickets. In these cases, even without considering spatial cloaking (which is not recommended), the attacker cannot trace the user and distinguish the sequence of user's queries and hence, his identity, path and activities. The proposed method satisfies the unlinkability metric and thus preserves the privacy of the user. Note that in the central trusted server approach, the system is vulnerable to these threats in the case of compromising the trusted server or gaining control over it by the intruder.

The final scenario is detecting user's speed in highways. Most of the previous methods do not preserve the user against this threat because they do not satisfy the unlinkability metric. Even knowing two consecutive queries is sufficient to detect over-speeding. In this case, by using cameras, the user is distinguishable by knowing the approximate location and using the pseudonym, spatial cloaking, and the fake path could not help. Since the only required information is two correlated queries, by using other methods, the user will be distinguished. A central trusted server is not completely trustable because it could be controlled by the government or the insurance companies. The proposed method is reliable in this case because in this model, there are not any correlated queries, and no one could determine that two queries belong to one user. Thus, over-speeding detection is impossible.  

\subsection{Attacks against $P^4QS$} 
In this part, we are going to explain the possible attacks against the $P^4QS$. It uses tickets and satisfies anonymity and unlinkability metrics and thus preserves user's  privacy. However, using tickets will cause the vulnerability in some situations as explained below.

\textbf{Non-cooperative nodes:}
It is possible that $P_{A_j}$ or $P_{B_m}$ may not cooperate with the users, and the user could not get its response during this period. In this case, it would be hard for the user to find the non-cooperative peer since he only knows that the response is not available. If the problem is with the $P_{B_m}$, it can be easily solved by sending another query. By having a new hash value, the new broker will be assigned to deliver the response. However, since the user is in a particular location, the $P_{A_j}$ is constant and thus, the system needs to determine the non-cooperative peers and remove them from the hash table. This could be done by announcing the non-delivery of the response. If these announcements exceed a specific threshold, the $P_{A_j}$ will be removed from DHT, but this threat remains in the proposed model.

\textbf{Sending used tickets to other peers:}
In the $P^4QS$, each peer exchanges its tickets with other trusted peers to avoid identification by the LBS through using specific tickets. In this case, a peer can have malicious activities such as sending invalid tickets or used tickets. Invalid tickets could be recognized through the identifiers as mentioned in Section~\ref{Protocol Execution}, but the used tickets cannot be determined. The only possible solution is that the server adds a time stamp to the tickets to show their valid time. Thus by limiting the usage time of the tickets, a malicious peer has does not have enough time to use and resend the tickets to other peers, but the model is still vulnerable against this attack.

\subsection{Experimental Evaluation} \label{experiments}
The $P^4QS$ is evaluated by implementing a server-based system and a proof-of-concept application on Android operating system to prove the feasibility and evaluate  the overall performance of the protocol. For this purpose, two experiments are performed: $(i)$  We first perform some tests on a local network (Android device, emulator, and server are all connected to a local, isolated network), $(ii)$ then the same tests are performed in the Internet.

In the implemented protocol, each user chooses a random pair in (-270, 270) interval, that is called $RA$. Note that the sum of each pair of (longitude, latitude) on the Earth is in this interval. Clients, either as real Android devices or virtual clients in the emulator, will form a DHT in the server and will be sorted based on their $RA$s. Also, each client has the pair ($RA$, $IP$ Address) of  four successors and four predecessors of itself in the DHT. The $N^{th}$ successor of client \textit{A} is $2^{(N-1)}$th node in the DHT after $A$ and $M^{th}$ predecessor of client $A$ is $2^{(M-1)}$th node in the DHT before $A$.

The implemented querying process consists of a few steps explained below:
First, the client needs to find its anonymizer. So he calculates the sum of its longitude and latitude and then finds the client whose \textit{RA} is closest to the calculated sum. After that, the client chooses one of its tickets to put it in the query and find the broker as well.

Tickets are created by the server and sent directly to clients (for example, in reply to a join message). For creating a ticket, the server chooses a random pair in (-270, 270) interval and attaches it to the known string "server" and finally, encrypts it with the server's private key. So, it is only the server that can create valid tickets.

After the client chooses a ticket, it decrypts the ticket with server's public key. Now, the client can find the broker of the query. The broker is the client with the nearest RA to the random number of the tickets. Now it is the time to create the query. Query is a message that consists of four strings:\\
1- Client's longitude\\
2- Client's latitude\\
3- (Query text + ticket + broker's IP address) encrypted with a symmetric key.\\
4- (The symmetric key) encrypted with the server's public key.\\
Then the query is sent to the anonymizer, and a request for the answer to this query is sent to the broker. The anonymizer waits for five seconds (or less if four queries have been received) to gather some queries. If there are less than four queries received, the anonymizer will create some queries to have four queries to anonymize. Now, the anonymizer changes the longitude and latitude of the queries to anonymize them. This process is a trade-off between the quality of service (accuracy of the answer) and privacy. The anonymized queries are sent to the server. The server decrypts the ticket of each query, checks the validity of the ticket and creates an answer message, and sends it to the broker responsible for that answer. When the broker receives an answer from the server, it decrypts the ticket, checks the validity of the ticket and then sends the answer back to the client who has requested for that answer.

\subsubsection{Local Evaluation}
\begin{figure}
\centering
\includegraphics[width=0.9\linewidth]{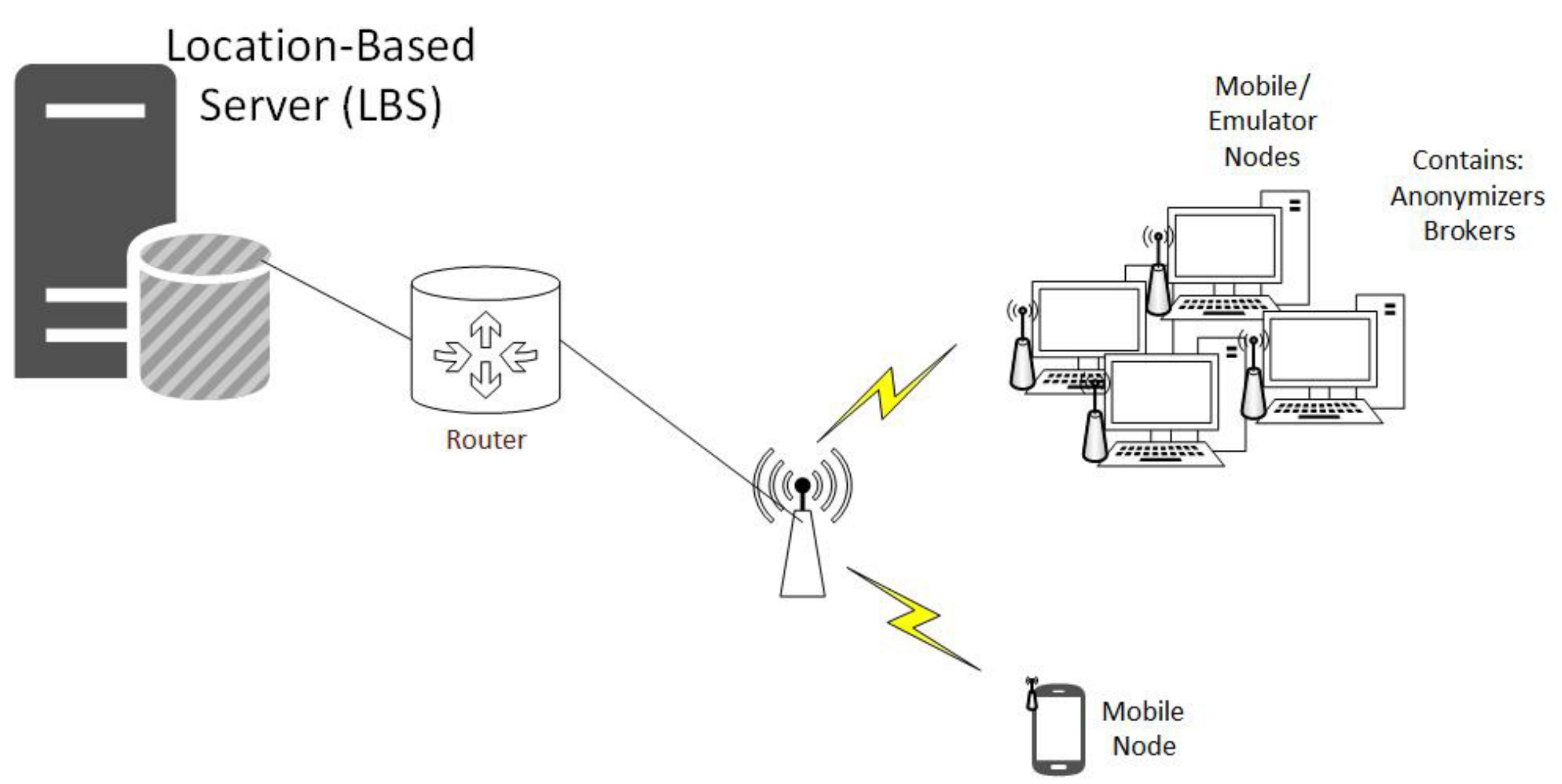}
\caption{Local Network Architecture}
\label{fig:Exp-local}
\end{figure}

In this experiment, the server was connected to an access point using an ethernet cable and the cell phone and emulator nodes were attached to the same access point using a wireless connection. (As shown in Fig.~\ref{fig:Exp-local} )

Hardware specifications of devices used in this experiment are as follows:\\
\textbf{Android device}: Samsung GT-I9000 mobile phone, which has 1GHz Cortex-A8 CPU, 512MB of RAM, 8GB of internal memory and Android version 4.4.4.\\
\textbf{Emulator}: A 2.6GHz Dual-core Intel Core i5 computer, with 8GB RAM and 256GB PCIe-based flash storage.\\
\textbf{Server}: A desktop computer with Dual-Core CPU E5300 @ 2.60GHz, x64-based processor and 2GB RAM.

In this experiment, two different approaches were taken. First, we ran Android application and the emulator as the ordinary clients. The Android device and virtual devices were used to send queries and receive the answer.

Table~\ref{tbl:combination mode local} shows the mean of the querying time (the time between sending the query to the anonymizer and receiving the answer from the broker) and the mean time of finding the  broker for the mobile node and emulator nodes. Wait time is a random time the client waits after receiving an answer. After that time, the client sends another query.

\begin{table*}[ht]
  \caption{Wait time for combination mode within local network}
  \begin{center}
  
      \begin{tabular}{ c | c | c |  c |  c }\toprule 
      
      \textbf{Time(ms)} / \textbf{Number of Nodes} & \textbf{20} & \textbf{40} & \textbf{60} & \textbf{80} \\ \rowcolor[gray]{.9} 
       \hline
      \emph{Receiving query mean time (emulator)}  & 4855.44 & 4993.72 & 5071.60 & 5802.11      
          \\ 
       \hline
      \emph{Finding broker mean time (emulator)}  & 10.98 & 14.03 & 9.76 & 5.21  \\ \rowcolor[gray]{.9}
       \hline
       \emph{Receiving query mean time (mobile)}  & 4354.41  & 4867.50 & 5199.06 & 5361.60 \\
       \hline
      \emph{Finding broker mean time (mobile)}  &  46.48 & 31.63 & 45.73 & 53.71 \\ 
        \hline              
      \end{tabular}
 \label{tbl:combination mode local}
  \end{center}
 \end{table*}

Moreover, a particular configuration was tested to analyze the abnormal situations. In this test, the anonymizer of all clients in the  emulator is set to be the real device. So the load effect of the protocol on a single node is analyzed. Table~\ref{tbl:local all load} shows the results of this test.

 \begin{table*}[ht]
  \captionof{table}{Sending all load on a single mobile node in local network}
  \begin{center}  
      \begin{tabular}{ p{6cm} | c | c | c}\toprule
      
      \centering{\textbf{Maximum wait time (ms)/number of emulator nodes}} & \textbf{20} & \textbf{50} & \textbf{100} \\ \rowcolor[gray]{.9}
       \hline
       \emph{10000} &678.42 & 1668.78 & 2947.87
          \\ 
       \hline
      \emph{20000} & 1492.32 & 1585.00 & 2712.63 \\ \rowcolor[gray]{.9} 
       \hline
      \emph{30000} & 1828.21  & 2272.16 & 2332.19 \\
       \hline
      \emph{40000} & 2156.14 & 2319.00 & 3122.87 \\\rowcolor[gray]{.9}
        \hline
        \emph{50000} & 2062.66 &  2486.10 & 3042.99  \\
          \hline
        \emph{60000} & 1954.89 & 3267.05 & 3019.08 \\
                  \hline           
      \end{tabular}
 \label{tbl:local all load}
  \end{center}
  \end{table*}

As the results indicate, the processing load of the proposed method is not significant for any user node over the network.

\subsubsection{Internet-based Evaluation}
\begin{figure}
\centering
\includegraphics[width=0.7\linewidth]{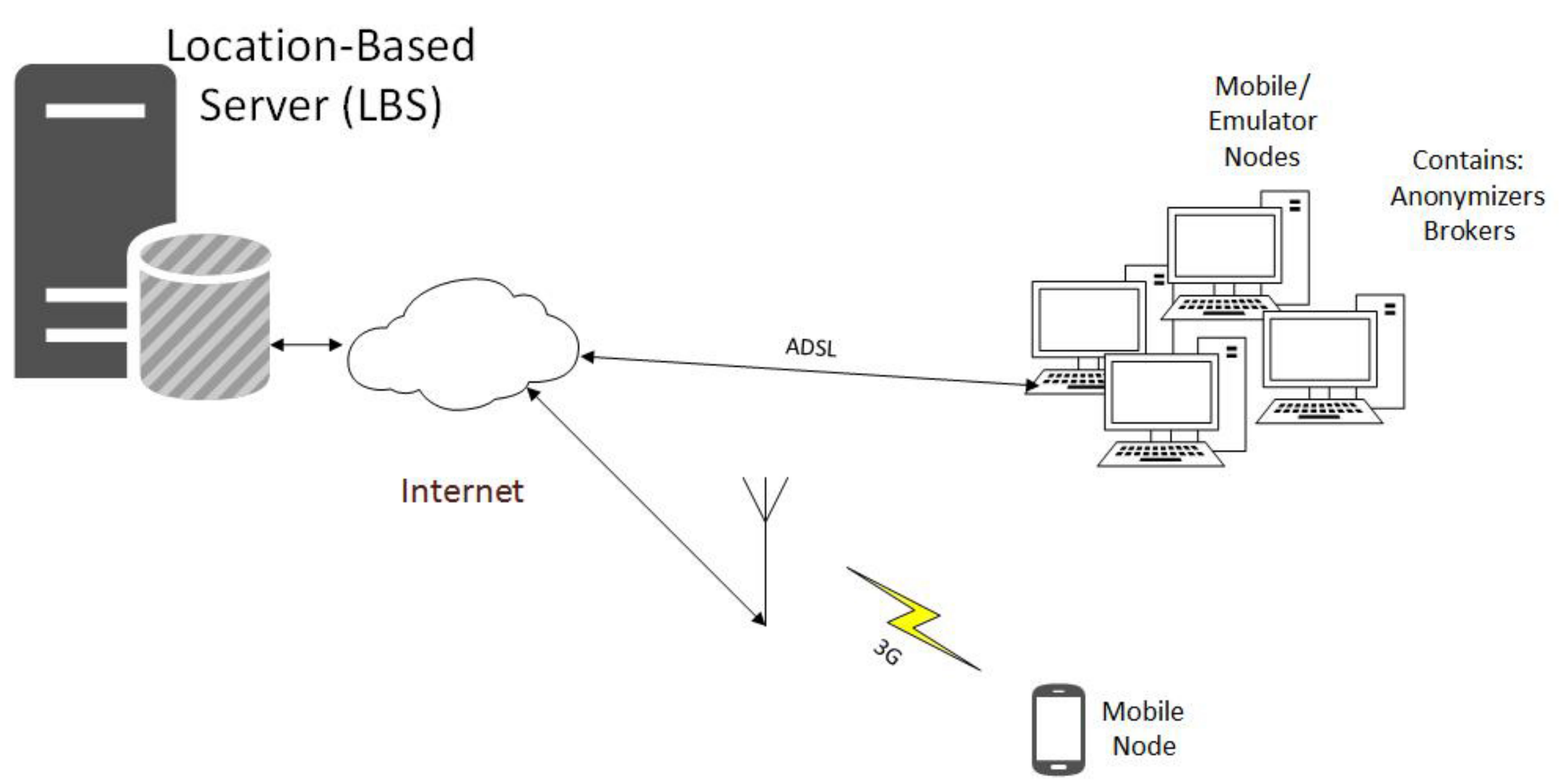}
\caption{Experimental architecture over the Internet}
\label{fig:Exp-int}
\end{figure}

In this part, the protocol was evaluated over the Internet using multiple devices with different connection types. The cell phone was connected to the Internet using a wireless ADSL modem. For the emulator and the server, two virtual private servers with valid IP addresses were used. So this test could be more close to the real-world infrastructures. (Fig.~\ref{fig:Exp-int})

Table~\ref{tbl:internet mixed} shows the mean of querying time over the Internet with different wait times before sending the next query.

\begin{table*}[ht]
  
  \begin{center}
  \captionof{table}{Wait time for combination mode over the Internet}
      \begin{tabular}{ c | c | c |  c |  c} \toprule
      
      \textbf{Time(ms) / Number of Nodes} & \textbf{20} & \textbf{40} & \textbf{60} & \textbf{80} \\ \rowcolor[gray]{.9} 
       \hline
      \emph{Receiving query mean time (emulator)}  & 7539.92 & 13548.82 & 15575.06 & 21280.23     
          \\ 
       \hline
      \emph{Finding broker mean time (emulator)}  & 3271.46 & 11686.93 & 13613.34 & 13197.20  \\ \rowcolor[gray]{.9}
       \hline
       \emph{Receiving query mean time (mobile)}  & 11876.76 & 14599.56 & 14238.75 & 14387.89 \\
       \hline
      \emph{Finding broker mean time (mobile)}  &  3285.76 & 5719.37 & 5975.78 & 5891.31 \\
        \hline              
      \end{tabular}
 \label{tbl:internet mixed}
  \end{center}
   \end{table*}

With a similar scenario, all virtual devices send their queries to one mobile anonymizer node. Table~\ref{tbl:internet all load} shows the mean of  the querying time in this scenario.

 \begin{table*}[ht]
  
  \begin{center}
  \captionof{table}{Sending all load on a single mobile node over the Internet}
      \begin{tabular}{ p{6cm} | c | c | c}\toprule
      
      \centering{\textbf{Maximum wait time(ms)/number of emulator nodes}} & \textbf{20} & \textbf{50} & \textbf{100} \\ \rowcolor[gray]{.9}
       \hline
       \emph{10000} & 4754.27 & 20948.73 & 55975.71           \\ 
       \hline
      \emph{20000} & 4084.68 & 15154.70 & 42778.03 \\  \rowcolor[gray]{.9}
       \hline
      \emph{30000} & 4648.85 & 19952.28 & 46668.33 \\
       \hline
      \emph{40000} & 4484.59 & 18305.65 & 48671.62 \\ \rowcolor[gray]{.9}
        \hline
        \emph{50000} & 5096.97 & 14300.29 & 43559.16  \\
          \hline
        \emph{60000} & 4830.29 & 12913.29 & 52732.65 \\ 
                  \hline           
      \end{tabular}
 \label{tbl:internet all load}
  \end{center}
   \end{table*}

All of these experiments indicate that the $P^4QS$ does not have a significant delay time or computational overhead. Thus, it can be easily deployed to achieve privacy with anonymization without any extra costs such as a trusted server.

\section{Conclusion}  \label{Cunclusion}
In this paper, we introduced a novel peer-to-peer architecture for anonymizing location-based queries. In the proposed architecture, each node was responsible for anonymizing a zone and at the same time, it could use location-based services as a client peer. By using appropriate symmetric and asymmetric encryptions, each node just has the required information. So it could not have any more details about the owner of the query or the sequence of her queries. The ticket exchange protocol supported the execution and protected against DoS attacks. The $P^4QS$ was implemented and tested on the local network and on the Internet. Experimental results showed that the $P^4QS$ had a negligible  effect on the time and computation costs.





%

\balance

\bibliographystyle{unsrt}
\bibliography{bare_conf}

\end{document}